\documentclass[10pt, final, conference, letterpaper, onecolumn, oneside]{./IEEEtran}

\usepackage{cite}
\usepackage{graphicx}
\usepackage{subfigure}
\usepackage{amsmath}
\usepackage{amssymb}
\usepackage{epsfig}
\usepackage{times}
\usepackage{color}
\usepackage{yhmath}

\newtheorem{pavikl}{\textbf{Lemma}}

\newtheorem{pavikp}{\textbf{Proposition}}

\begin{document}

\title{Worst-case Compressibility of Discrete and Finite Distributions}%
\author{\IEEEauthorblockN{Samar Agnihotri\IEEEauthorrefmark{2} and Rajesh Venkatachalapathy\IEEEauthorrefmark{4}}%
\IEEEauthorblockA{\IEEEauthorrefmark{2}CEDT, Indian Institute of Science, Bangalore - 560012, India\\}
\IEEEauthorblockA{\IEEEauthorrefmark{4}Systems Science Graduate Program, Portland State University, Portland, OR 97207\\}
Email: samar@cedt.iisc.ernet.in, venkatr@pdx.edu%
}

\IEEEspecialpapernotice{(Extended Abstract)}

\maketitle

\begin{abstract}
In the worst-case \textit{distributed source coding} (DSC) problem of \cite{allerton08}, the smaller cardinality of the support-set describing the correlation in informant data, may neither imply that fewer informant bits are required nor that fewer informants need to be queried, to finish the data-gathering at the sink. It is important to formally address these observations for two reasons: first, to develop good worst-case information measures and second, to perform meaningful worst-case information-theoretic analysis of various distributed data-gathering problems. Towards this goal, we introduce the notions of \textit{bit-compressibility} and \textit{informant-compressibility} of support-sets. We consider DSC and distributed function computation problems and provide results on computing the bit- and informant- compressibilities regions of the support-sets as a function of their cardinality.
\end{abstract}

\section{Introduction}
\label{sec:Intro}
Let us consider a data-gathering sensor network, where a sink node is interested in losslessly gathering the data from $N$ sensor nodes. Sensor data in such networks is assumed to be correlated. For the purpose of data-gathering at the sink, the sensor nodes communicate with the sink and communication consumes energy. However, sensor nodes are assumed to be severely energy-constrained. Therefore, to achieve long operational lifetime of the network, we need to come up with communication protocols and architectures which allow the sensor nodes to expend as little energy in communication as possible. One approach to accomplish this is to use the framework of \textit{distributed source coding} (DSC), \cite{104xiongLiveris, 104chouPetrovic, 105adler}.

Motivated by the problem of maximizing the worst-case operational lifetime of the data-gathering sensor networks, in \cite{allerton08} we addressed a variant of DSC problem. We assumed that the correlation in sensor data is modeled by discrete and finite probability distribution $\cal P$, as in \cite{104chouPetrovic, 105adler}. Further for the worst-case analysis, we introduced a new information measure, called \textit{information ambiguity}. We assumed asymmetric communication scenario \cite{097kushilevitzNisan}, where only the sink knows $\cal P$. Finally, we gave a communication protocol (\textbf{bSerCom} protocol) that computed the optimal number of informant bits required, in the worst-case, to let the sink learn the data-vector revealed to the informants when the data-vectors were derived from $\cal P$.

While generalizing some portion of our work in \cite{allerton08}, we made following interesting observations:

\textit{Observation 1:} In general, the cardinality of a support-set is a good indicator of the minimum number of informant bits required, in the worst-case, to describe any data-vector derived from it. Our observation indicated some violations of this general correspondence. There are situations, where to describe the data-vectors derived from two support-sets with same cardinality, different minimum number of informant bits are required. Similarly, there are instances where to describe the data-vectors derived from a support-set with smaller cardinality requires more number of informant bits than those derived from a support-set with larger cardinality. This implies that the worst-case information measures such as information ambiguity \cite{isita08}, which are based only on the cardinality of the support-set, cannot be reliably used to predict the number of informant required, in \textit{all} scenarios.

\textit{Observation 2:} In our work on the application of the worst-case DSC to the problem of maximizing the worst-case operational lifetime of data-gathering sensor networks, we encountered the situations where one or more nodes were much more energy-constrained than other nodes. Therefore, for the long lifetime of network, querying such energy-starved nodes needed to be avoided. However, there exist support-sets describing the correlation in sensor data, which have small cardinality, yet require querying all sensor nodes to let the sink to unambiguously learn of any data-vector derived from such support-sets, even if it leads to smaller network lifetime. This motivated us to address in the DSC framework, not only the problem of minimizing the number of informant bits, but also the problem of minimizing the number of informants to be queried to allow the sink to exactly learn the data-vector revealed to the informants.

It is important to formally address these two observations because first, it leads to the development of \textit{good} worst-case information measures and second, it allows for sound worst-case analysis of various distributed data-gathering problems of interest. This paper documents our efforts in this direction and some insights we gained from such efforts.

\textit{Related work:} The work on the application of \textit{compressed sensing} (\cite{106donoho, 106candes}) framework to distributed source coding problem, as in \cite{106duarteWakinBaraniuk, 108hauptBajwaNowak}, relates most closely to our work, though more in terms of the problem being addressed and some of the results obtained, than in terms of the approach taken to solve the problem.

\section{Worst-case Compressibility of a Discrete and Finite Distribution}
\label{sec:sparsity}
Let us consider a discrete and finite distribution $\cal P$ for $N$ random variables $X_i, i \in \{1, \ldots, N\}, X_i \in {\cal X}$, where $\cal X$ is the discrete and finite alphabet of size $|{\cal X}|$.

Let us consider a $N$-tuple of random variables $(X_1, \ldots, X_N) \sim {\cal P} = p(x_1, \ldots, x_N)$. The \textit{support set} of $(X_1, \ldots, X_N)$ is defined as:
\begin{equation}
\label{eqn:supp_set_dstrbnN}
S_{X_1, \ldots, X_N} \stackrel{\textrm{def}}{=} \{(x_1, \ldots, x_N) | p(x_1, \ldots, x_N) > 0 \}
\end{equation}
Let us denote the cardinality of $S_{X_1, \ldots, X_N}$ as $\mu_{X_1, \ldots, X_N}, \mu_{X_1, \ldots, X_N} = |S_{X_1, \ldots, X_N}|$. In \cite{allerton08, isita08}, we address $\mu_{X_1, \ldots, X_N}$ also as the \textit{information ambiguity} of $(X_1, \ldots, X_N)$. So, the minimum number of bits required to describe an element of $S_{X_1, \ldots, X_N}$, in the worst-case, is $\lceil \log \mu_{X_1, \ldots, X_N} \rceil$. Note that all the logarithms used in this paper are to the base 2.

The \textit{support set} $S_{X_i}$ of $X_i, i \in \{1, \ldots, N\}$, is the set
\begin{equation}
\label{eqn:supp_set_varN}
S_{X_i} \stackrel{\textrm{def}}{=} \{x_i: \mbox{ for some } x_{-i}, (x_{-i}, x_i) \in S_{X_1, \ldots, X_N}\}, \mbox{with } x_{-i} \stackrel{\textrm{def}}{=} \{x_1, \ldots, x_N\} \setminus x_i
\end{equation}
of all possible $X_i$ values. Let us denote the cardinality of $S_{X_i}$ as $\mu_{X_i}, \mu_{X_i} = |S_{X_i}|$. So, the minimum number of bits required to describe an element of $S_{X_i}$, in the worst-case, is $\lceil \log \mu_{X_i} \rceil$.

We define \textbf{sparsity} $\gamma$ of distribution $\cal P$ as the fraction of $|{\cal X}|^N$ (the maximum cardinality of the support-set of any discrete and finite distribution $\cal P$ of $N$ random variables which derive their values from a discrete alphabet of size $|\cal X|$), that is:
\begin{equation}
\label{eqn:sparsity_def}
\gamma = \frac{\mu_{X_1, \ldots, X_N}}{|{\cal X}|^N}, \gamma \in \; ]0, 1]
\end{equation}

Let $\#_b$ denote the minimum number of informant bits required, in the worst-case, to describe any data-vector derived from the distribution $\cal P$. Similarly, let $\#_n$ denote the minimum number of informants required to be queried, in the worst-case, to describe any data-vector derived from the distribution $\cal P$. Let us introduce
\begin{equation}
\label{eqn:betaDef}
\beta = \frac{\#_b}{\sum_{i = 1}^N \lceil \log \mu_{X_i} \rceil} \:,  \quad \frac{\lceil \log \mu_{X_1, \ldots, X_N} \rceil}{\sum_{i = 1}^N \lceil \log \mu_{X_i} \rceil} \le \beta \le 1
\end{equation}
and
\begin{equation}
\label{eqn:etaDef}
\eta = \frac{\#_n}{N} \:, \quad \frac{1}{N} \le \eta \le 1
\end{equation}

Then, the \textbf{worst-case compressibility} of distribution $\cal P$ is defined as one or both of the following:

\textbf{Bit-compressibility:} a discrete and finite distribution $\cal P$ is called bit-compressible if the minimum number of informant bits required, in the worst-case, to describe any input-vector derived from the distribution is smaller than the total number of informant bits required to describe an instance of $(X_1, \ldots, X_N)$ when each $X_i$ is separately described in $\lceil \log \mu_{X_i} \rceil$ bits, that is, if for the given distribution $\beta < 1$. Otherwise if $\beta = 1$, it is called bit-incompressible.

\textbf{Informant-compressibility:} a discrete and finite distribution $\cal P$ is called informant-compressible if the minimum number of informants required to participate in the worst-case data-gathering is smaller than the total number of informants, that is, if for the given distribution $\eta < 1$. Otherwise if $\eta = 1$, it is called informant-incompressible.

Sparsity $\gamma$ of a distribution generally implies its bit-compressibility. That is, for a sample space defined by the pair $(|{\cal X}|, N)$, we have following sequence of implications:
\begin{equation*}
\gamma \rightarrow 0 \implies \mu_{X_1, \ldots, X_N} \rightarrow 1 \implies \lceil \log \mu_{X_1, \ldots, X_N} \rceil \rightarrow 0
\end{equation*}
In other words, the data-vectors derived from sparse distributions require fewer informant bits to describe than those derived from dense distributions. However, we needed to introduce the notion of bit-compressibility of a distribution to address the scenarios where it is not so, that is, the discrete and finite distributions which are sparse, but incompressible.

Similarly, the sparsity of a distribution implies its informant-compressibility, though the implication in this case is not so obvious. In this case too, there are sparse distributions which require more informants to participate in data-gathering than dense distributions and there are sparse distributions which are informant-incompressible. To formally address such scenarios, we needed to introduce the notion of informant-compressibility of the distribution. As mentioned in the Introduction, the notion of informant-compressibility is quite important in data-gathering sensor networks where the sink is interested in exactly learning the instance of data-vector revealed to the sensors, without querying one or more of energy-constrained sensor nodes and we want to know whether it can be done for the given support-set. Another implication of informant-compressibility in the context of sensor networks is the following. If the distribution describing the correlation in sensor data is informant-compressible, then the sink needs to query only $K, K < N$, sensor nodes, which implies that only these $K$ nodes need to sample the phenomenon being observed. Therefore, only $K$ observations are sufficient to exactly learn the sample values at all $N$ sensor locations when $N - K$ sensors are not even sensing. In terms of energy costs, this not only leads to saving of the communication energy, but also the energy expended in sensing and processing at the concerned nodes. Further, this also allows the nodes to more equitably share the burden of data-gathering as any $K$ out of $N$ nodes need to be active at any time, so in $N$ successive rounds, a node is active only during $K$ rounds.

\subsection{Illustrations of breakdown of sparsity-compressibility relation}
\label{subsec:exas}

\begin{figure}[!t]
\centering
\includegraphics[width=7.0in]{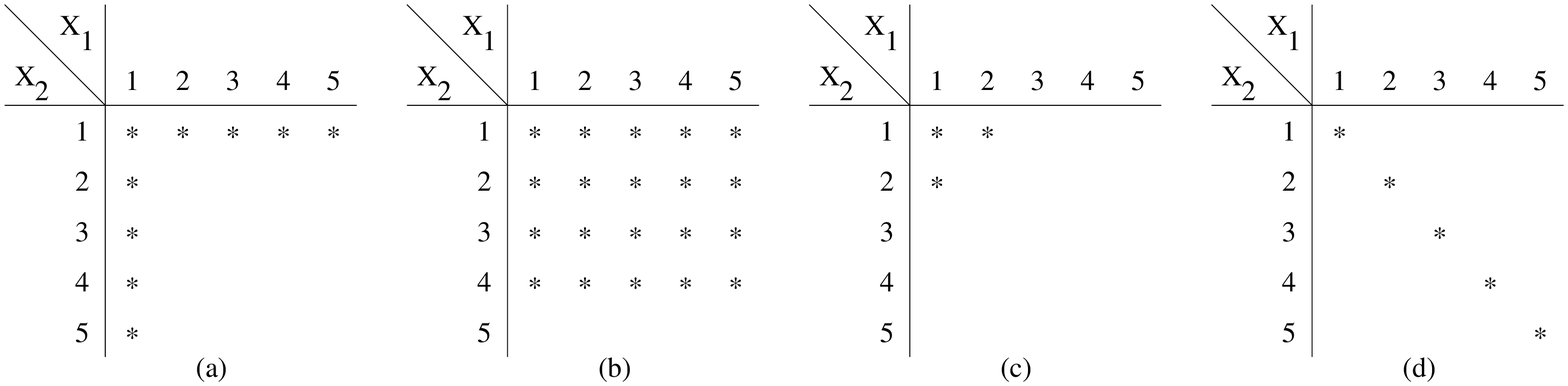}
\vspace{-0.1in}
\caption{Various support-sets for sparsity versus compressibility}
\label{fig:cardinalityExa}
\vspace{-0.25in}
\end{figure}

\textit{Example 1:} Let us first address an example of situations where to describe the data-vectors derived from the support-set with smaller cardinality actually requires more number of informant bits, in the worst-case, than to describe the data-vectors derived from the support-set with larger cardinality. Let us consider distributions in figures~\ref{fig:cardinalityExa}.(a)-(b). The support-set in (a) has cardinality $9$, but it requires $6 (= 3 + 3)$ informant bits, in the worst-case, to describe any data-vector. On the other hand, the support-set in (b) has cardinality of $20$, yet it requires $5 (= 2 + 3)$ informant bits, in the worst-case, to describe any data-vector.

\textit{Example 2:} Now, let us discuss an example of situations where to describe the data-vectors derived from the support-set with smaller cardinality actually requires more number of informant to be queried, in the worst-case, than to describe the data-vectors derived from the support-set with larger cardinality. Consider the distributions in figures~\ref{fig:cardinalityExa}.(c)-(d). The support-set in (c) has cardinality $3$, but it requires both the informants to communicate to describe any data-vector. On the other hand, the support-set in (d) has cardinality $5$, yet it requires only one informant to communicate to describe any data-vector.

These two examples demonstrate that sparser\footnote{Note that the sparsity of a distribution, as defined in \eqref{eqn:sparsity_def}, is a stronger notion than the cardinality of the corresponding support-set, as it allows us to discuss on the same footing even the distributions derived from different sample-spaces defined by different $(|{\cal X}|, N)$-pairs. However, when all the distributions under consideration are derived from same sample-space defined by given $(|{\cal X}|, N)$-pair, as in this paper, then it is equivalent to talk in terms of sparsity and cardinality of the distributions and we follow this in the rest of the paper.} distributions may neither be more bit-compressible nor be more informant-compressible. Sparsity of a distribution may not imply its compressibility in either sense, defined above. Also, it should be noted that this holds for any communication protocols, consistently deploying any, \textit{a priori} agreed upon, encoding and decoding scheme over all support-sets in the sample space.

In the next two subsections, we address the sparsity-compressibility issue in more detail for two distributed communication scenarios. First, the distributed source coding problem in asymmetric communication scenarios, as in \cite{allerton08} and second, the distributed function computation problem in asymmetric communication scenarios, as in \cite{paper_1}.

\subsection{Sparsity versus compressibility for DSC}
\label{subsec:forDSC}
We first discuss the sparsity-compressibility issue with respect to the DSC problem in asymmetric communication scenarios of \cite{allerton08}. We address both, bit-compressibility and informant-compressibility. We are interested in answering two questions: (1) ``what is the smallest cardinality of support-sets for which there is at least one support-set that is incompressible?'', and (2) ``what is smallest cardinality of support-sets, such that all support-sets with cardinalities more than this are incompressible?''. These questions are addressed in following lemmas, which we state without proof for the sake of brevity.

\underline{For bit-compressibility:}

\begin{pavikl}
\label{lemma:m1DSC}
The smallest cardinality $M_1$ of the support-sets for which there is at least one support-set that requires each of $N$ informants to send (worst-case) $\lceil \log |{\cal X}| \rceil$ bits is $M_1 = N(|{\cal X}|-1) + 1$.
\end{pavikl}

\begin{pavikl}
\label{lemma:m2DSC}
The smallest cardinality $M_2$ of the support-sets, such that for all support-sets with cardinality $m, m > M_2$, each of $N$ informants sends (worst-case) $\lceil \log |{\cal X}| \rceil$ bits is $M_2 = |{\cal X}|^{N-1}(|{\cal X}|-1)$.
\end{pavikl}

Similarly, \underline{for informant-compressibility:}

\begin{pavikl}
\label{lemma:m3DSC}
The smallest cardinality $M_3$ of the support-sets for which there is at least one support-set that requires each of $N$ informants to transmit is $M_3 = N+1$.
\end{pavikl}

\begin{pavikl}
\label{lemma:m4DSC}
The smallest cardinality $M_4$ of the support-sets, such that for all support-sets with cardinality $m, m > M_4$, each of $N$ informants transmits is $M_4 = |{\cal X}|^{N-1}$.
\end{pavikl}

The support-sets in Figure~\ref{fig:cardinalityExa}.(a)-(d) correspond to the support-sets with cardinalities $M_1, M_2, M_3$, and $M_3$ of Lemma~\ref{lemma:m1DSC}-Lemma~\ref{lemma:m4DSC}, respectively, for $|{\cal X}| = 5$ and $N = 2$.

The total number of support-sets with cardinality $M_1$ are $\binom{|{\cal X}|^N}{N(|{\cal X}|-1) + 1}$. Out of these, the number of support-sets which result in every informant sending its worst-case number of bits (bit-incompressible) is $|{\cal X}|^{N-1} ((N-1)(|{\cal X}| - 1) + 1)$. Therefore, the fraction of all possible support-sets with cardinality $M_1$ which are bit-incompressible is
\begin{equation}
\label{eqn:m1ratio}
\frac{|{\cal X}|^{N-1} ((N-1)(|{\cal X}| - 1) + 1)}{\binom{|{\cal X}|^N}{N(|{\cal X}|-1) + 1}}
\end{equation}

Similarly, the total number of support-sets with cardinality $M_3$ are $\binom{|{\cal X}|^N}{N + 1}$. Out of these, the number of support-sets which result in every informant participating in data-gathering (informant-incompressible) is $|{\cal X}|^N (|{\cal X}| - 1)^N$. Therefore, the fraction of all possible support-sets with cardinality $M_3$ which are informant-incompressible is
\begin{equation}
\label{eqn:m3ratio}
\frac{|{\cal X}|^N (|{\cal X}| - 1)^N}{\binom{|{\cal X}|^N}{N + 1}}
\end{equation}

For large values of $N$ and $|{\cal X}|$, it is easy to see, using Stirling's approximation for factorial, that the ratios in \eqref{eqn:m1ratio} and \eqref{eqn:m3ratio} are much smaller than $1$. Similar ratios can be computed for the numbers of bit- and informant- incompressible support-sets corresponding to cardinalities $M_2$ and $M_4$, respectively, and shown to be very close to $1$. This implies that most of the support-sets with small cardinality (\textit{sparse} support-sets) are compressible and most of the support-sets with large cardinality (\textit{dense} support-sets) are incompressible, confirming our intuition regarding such situations.

\begin{figure}[!t]
\centering
\includegraphics[width=7.0in]{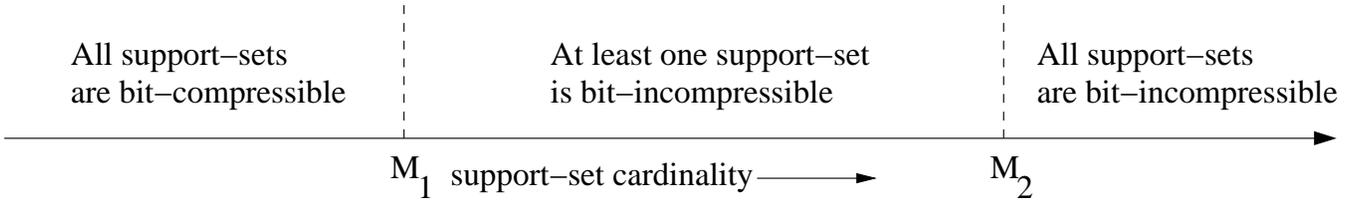}
\vspace{-0.1in}
\caption{Bit-compressibility versus cardinality of support-sets}
\label{fig:bitCompressibilityRegions}
\vspace{-0.1in}
\end{figure}

\begin{figure}[!t]
\centering
\includegraphics[width=7.0in]{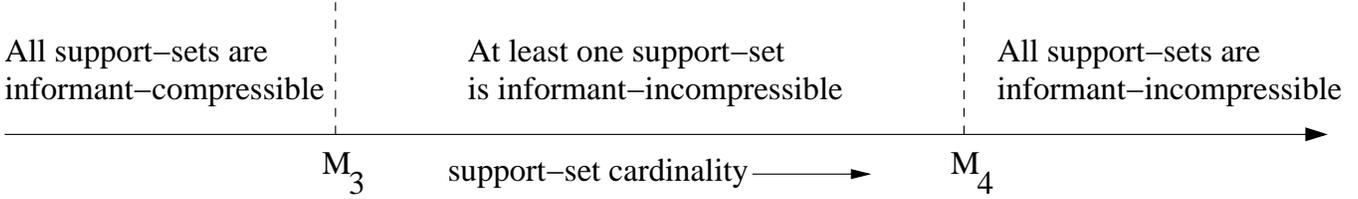}
\vspace{-0.1in}
\caption{Informant-compressibility versus cardinality of support-sets}
\label{fig:informantCompressibilityRegions}
\vspace{-0.25in}
\end{figure}

In figures~\ref{fig:bitCompressibilityRegions} and \ref{fig:informantCompressibilityRegions}, we plot the bit-compressibility and informant-compressibility regions, respectively, with respect to the cardinality of the support-set. It should be noted from these compressibility-region plots that all distributions with cardinalities smaller than $M_3$ are both, bit- and informant- compressible, while all distributions with cardinalities more than $M_2$, are both bit- and informant- incompressible.

Further, comparing the values of $M_1, M_2, M_3$ and $M_4$ defined in Lemma~\ref{lemma:m1DSC}-Lemma~\ref{lemma:m4DSC}, we have:
\begin{pavikp}
$M_3 < M_1 < M_4 < M_2$, for all $N > 2$ and $|{\cal X}| > 2$.
\end{pavikp}
This implies that there are informant-incompressible distributions, which are bit-compressible (for example,  the distributions with cardinality $m$ such that $M_4 \le m \le M_2$), but no bit-incompressible distribution that is informant-compressible.

\subsection{Sparsity versus compressibility for `bitwise OR'}
\label{subsec:forBOR}
We discuss here the sparsity-compressibility issue with respect to the distributed function computation problem in asymmetric communication scenarios as in \cite{paper_1}, with the function under consideration being `bitwise OR' function. As before, we are interested in both, bit-compressibility and informant-compressibility and address the questions on smallest cardinalities which lead to incompressibility. These questions are addressed in following lemmas, stated without proof for the sake of brevity.

\underline{For bit-compressibility:}

\begin{pavikl}
\label{lemma:m1bOR}
The smallest cardinality $M_1$ of the support-sets for which there is at least one support-set that for computing `bitwise OR' requires each of $N$ informants to send (worst-case) $\lceil \log |{\cal X}| \rceil$ bits is $M_1 = N(|{\cal X}|-1) + 1$ and there is only one such support-set.
\end{pavikl}

Similarly, \underline{for informant-compressibility:}

\begin{pavikl}
\label{lemma:m3bOR}
The smallest cardinality $M_3$ of the support-sets for which there is at least one support-set that for computing `bitwise OR' requires each of $N$ informants to transmit is $M_3 = N+1$.
\end{pavikl}

The number of support-sets with cardinality $M_3$, which require each informant to transmit, for computing `bitwise OR' is:
\begin{equation*}
\sum_{\stackrel{(i_1, \ldots, i_N)}{= (1, \ldots, 1)}}^{(|{\cal X}|, \ldots, |{\cal X}|)} \sum_{\stackrel{(r_1, \ldots, r_N)}{= (r_{i_1}^l, \ldots, r_{i_N}^l)}}^{(r_{i_1}^h, \ldots, r_{i_N}^h)} \prod_{j = 1}^{N} n_{i_j}^{r_j}
\end{equation*}
where $n_{i_j}^{r_j}$ is the number $r_j$ of equal output values of the `bitwise OR' function computation, different from its output value for $(i_1, \ldots, i_N)$ location, obtained by holding indices $\{i_1, \ldots, i_{j-1}, i_{j+1}, \ldots, i_N\}$ constant.

\subsection{Discussion}
\label{subsec:discussion}
There are two interesting inferences which can be drawn from our work on sparsity-compressibility issue.

\textbf{Worst-case compressibility is a stronger notion than sparsity:} A pessimistic interpretation says that for cardinalities as small as, $M_1$ and $M_3$, there are support-sets which require, respectively, each informant to send all bits in the representation of its data-value (bit-incompressibility) and each informant to transmit (informant-incompressibility). For such support-sets, interactive communication between the sink and informants is of no help in reducing the incompressibility of either type associated with the distribution. While an optimistic outlook can say that all support-sets with cardinalities smaller than $M_1$ are bit-compressible and all support-sets with cardinalities smaller than $M_3$ are informant-compressible. Further, even for the cardinalities as large as $M_2$ and $M_4$, there are bit- and informant- compressible support-sets, respectively. For such support-sets, the interaction between the sink and informants does help in reducing the number of informant bits sent and the number of informants needed to be queried, respectively. Also, when a support-set is bit-incompressible (such as support-set with cardinality more than $M_2$), then this simplifies the communication problem to trivial-case: sink asks each informant to send all of its information bits and the order in which informants communicate with the sink does not matter, leading to trivially simple communication protocols.

This along with the figures~\ref{fig:bitCompressibilityRegions}-\ref{fig:informantCompressibilityRegions}, leads us to assert that for finite and discrete distributions and with our definitions of sparsity and worst-case compressibility, the sparsity of the distribution is not a good indicator of its worst-case compressibility, in general. However, the worst-case compressibility of a distribution always implies its sparsity.

\textbf{Information ambiguity does not always indicate the worst-case compressibility:} Our work shows the inadequacy of the worst-case information measure of \textit{information ambiguity} \cite{allerton08, isita08}, which is defined as the cardinality of the support-set of a distribution, in indicating the worst-case compressibility. This is unlike the behavior of the average-case information measure of information entropy that has one-to-one correspondence with the average-case compressibility of a distribution. We argue that characterizing the support-set of a distribution, in the worst-case, by a single number, such as information ambiguity, involves discarding a lot of potentially useful information about how the support-set is populated. It is an interesting problem to develop a worst-case information measure that also accounts for such information embedded in a support-set.

\section{Conclusions and Future Work}
\label{sec:conclusions}
This paper documents our efforts towards formally addressing some observations with respect to worst-case distributed source coding problem of \cite{allerton08} and its applications to data-gethering sensor networks. We observed that in the distributed data-gathering problem in asymmetric communication scenarios, smaller cardinality of a support-set may neither lead to fewer number of informant bits nor to fewer number of informants to be queried. In this paper, we proposed two notions of \textit{bit-compressibility} and \textit{informant-compressibility} of support-sets to address these two situations, respectively. We provided results to compute the bit- and informant- compressibilities regions of the support-sets as a function of their cardinality. Our results led us to make the proposition that the worst-case compressibility of a support-set implies its sparsity, but converse may not always be true.

Our work on the interesting formal connections of the results presented here with the results obtained using compressed sensing framework for DSC is in a preliminary stage and has to be left for the future. Also, as we discussed above, it is interesting to develop a more powerful information measure for the worst-case information-theoretic analysis than information ambiguity and we are presently working at it.

\end{document}